\documentclass[conference]{IEEEtran}

\usepackage{cite}      

\usepackage{graphicx}  

\usepackage{amsmath}   

\usepackage{array}
\begin{document}


\title{QUANTIZED FUSION RULES FOR ENERGY-BASED DISTRIBUTED DETECTION IN WIRELESS SENSOR NETWORKS}



 
\author{\IEEEauthorblockN{Edmond Nurellari$^{1}$,  Sami Aldalahmeh$^2$,  Mounir Ghogho$^{1, 3}$ and Des McLernon$^1$}
\IEEEauthorblockA{$^1$University of Leeds, UK \\ $^2$Al-Zaytoonah University of Jordan, Jordan \\
$^3$International University of Rabbat, Morocco\\ elen@leeds.ac.uk, s.aldalahmeh@zuj.edu.jo, m.ghogho@ieee.org, d.c.mclernon@leeds.ac.uk}
}

\def\x{{\mathbf x}}
\def\L{{\cal L}}

\maketitle

\begin{abstract}
We consider the problem of soft decision fusion in a bandwidth-constrained wireless sensor network (WSN). The WSN is tasked with the detection of an intruder transmitting an unknown signal over a fading channel. A binary hypothesis testing is performed using the soft decision of the sensor nodes (SNs). Using the likelihood ratio test, the optimal soft fusion rule at the fusion center (FC) has been shown to be the weighted distance from the soft decision mean under the null hypothesis. But as the optimal rule requires a-priori knowledge that is difficult to attain in practice, suboptimal fusion rules are proposed that are realizable in practice. We show how the effect of quantizing the test statistic can be mitigated by increasing the number of SN samples, i.e., bandwidth can be traded off against increased latency. The optimal power and bit allocation for the WSN is also derived. Simulation results show that SNs with good channels are allocated more bits, while SNs with poor channels are censored.
\end{abstract}


%

\section{Introduction}
\label{sec:intro}
Distributed detection has been attracting significant interest in the context of WSNs \cite{Chamnerland2007} and \cite{les82}. This is due to the flexibility of WSNs, which can be seamlessly deployed over a wide geographic area for military monitoring and surveillance purpose \cite{Chen2006}. However, WSNs suffer from constrained bandwidth and limited on-board power. This poses challenges in the design of distributed detection algorithms, especially when the intruder's signature is unknown to the WSN. The main issue is to improve the detection by fusing the measurements provided by various SNs  in a manner that efficiently utilizes the scarce bandwidth and overcomes the limitations of a fading wireless channel.

The problem of decentralized detection in bandwidth constrained sensor networks has been addressed in \cite{Chamnerland2003}, where the authors investigated the design of sensor messages sent to the FC that minimize the error probability. The problem of detecting a known deterministic parameter is investigated in \cite{Xiao2005} under restricted channel capacity. The channel fading effect on distributed detection was tackled in \cite{Chenn2006}. In \cite{Barbarossa2013}, the authors addressed both issues of limited bandwidth and channel imperfections. They optimized the transmission power, which consequently dictated the number of allocated bits, for the detection of a known signal.

\indent In this paper, we consider the detection of an unknown signal, which is the case in many WSN applications. We find the optimal fusion rule for energy-based soft decision, through the use of the likelihood ratio test. However, it turns out that this rule is difficult to implement in practice. So we will suggest realizable suboptimal fusion rules that weight the soft decisions of the SNs based on the measurement's quality. Furthermore, a linear optimal fusion rule is derived that can serve when the probability distributions of the soft decisions are not known a-priori. Similarly, a simpler fusion rule is proposed based on the linear rule. Then, the previous algorithms are revisited under noisy, flat fading channels with limited bandwidth. Finally, the SN's transmitted powers are optimized to achieve the best probability of detection.

Section \ref{sec:System Model} presents the system model. Soft decision fusion rules are proposed in Section \ref{sec:Soft Decision Fusion Rules}. The quantization effect is discussed in Section \ref{sec:Quantized Soft Decision Fusion} and the optimal power allocation is derived in Section \ref{sec:Optimum sensor transmit power allocation}. Simulation results are given in Section \ref{sec:Simulation  Results} and  conclusions are presented in Section \ref{sec:conclusion}.
\section{System Model}
\label{sec:System Model}
\indent Consider a WSN with $M$ sensor nodes reporting to a FC tasked with the detection of any intruders. The intruder leaves a signature signal that is unknown to the WSN but it is assumed to be deterministic.  The $i^{th}$ SN collects $N$ samples that are corrupted by additive white Gaussian noise (AWGN) with zero mean and known variance $\sigma^2_{i}$. So, the measured signal takes one of the following forms, depending on the underlying hypothesis:
%
%
\begin{eqnarray}
\hspace*{-0.3cm}
\mathcal{ H}_0: x_i\left(n\right) &=&w_i\left(n\right) \\
\mathcal{ H}_1: x_i\left(n\right) &=& s_i\left(n\right)+w_i\left(n\right)
\end{eqnarray}
%
%
where $ n = 1, 2, \ldots, N$, $ i = 1, 2 \ldots, M$, $x_i(n)$ is the $n^{th}$ sample of the measured signal at the $i^{th}$ sensor, $s_i(n)$ is the intruder's signature signal and $w_i(n)\sim \mathcal{N}(0,{\sigma_{i}}^{2})$ is the AWGN. Furthermore, the noise samples are assumed to be identically and independently distributed (iid) across time and space. \\
\indent For optimal detection, the SNs should send their measurements to the FC, where the ultimate detection decision about the  intruder's presence will be made. However, this approach is not always feasible in the context of WSNs due to the limited bandwidth available. Thus, the WSN adopts a distributed detection algorithm in which the SNs send their quantized soft decisions (i.e., the quantized local test statistics) to the FC, which combines them to arrive at the global decision. Since the intruder's signal is unknown at the SNs, the optimal detector in this case would be the energy detector, which is implemented at the $i^{th}$ SN as follows:
%
%
\begin{equation}\label{eq:T_i}
     T_i=\sum \limits_{n=1}^N  \left| x_i \left( n \right) \right|^{2}.
\end{equation}
%
%
\indent The local soft test statistic $T_i$ is then quantized with $L_i$ bits and transmitted to the FC with power $p_i$  over a wireless channel. The channel suffers from zero mean AWGN with a variance of $\zeta_i$. Moreover, the wireless channel between the $i^{th}$ SN and the FC experiences flat fading with a channel gain $h_i$  (also assumed to be iid). 
The number of quantization bits at the  $i^{th}$ SN must satisfiy the channel capacity constraint:
%
%
\begin{equation}\label{eq:capacity}
     L_i\leq\frac{1}{2}\log_2\left(1+\frac{p_i h_i^{2}}{\zeta_i}\right) \mathrm{bits}, \ i = 1, 2, \ldots, M.
\end{equation}
%
%
\indent We will assume that the maximum channel capacity is utilized by the SNs. So our objective is to find the best soft fusion rule first, and then optimize the allocated power to maximize the detection probability.
%
%
\section{Soft Decision Fusion Rules}
\label{sec:Soft Decision Fusion Rules}
In this section, the optimal soft decision fusion rule is investigated given infinite bandwidth for each WSN, i.e., no quantization is required. However, it turns out that the optimal rule requires prior information about the signal's energy, which cannot be known in practice. Hence, suboptimal rules are proposed as an implementable alternative.

\subsection{Optimal Fusion Rule}
Given the local soft test statistic defined in (\ref{eq:T_i}), the optimal fusion rule follows from the likelihood ratio test:
%
%
\begin{equation}\label{eq:LRT}
  \mathrm{ LRT}\left(\boldsymbol  { T}\right) =\frac{p\left\{{ T_1}, { T_2}, ..., { T_M|\mathcal{ H}_1}\right\}}{p\left\{{ T_1}, {T_2}, ..., { T_M|\mathcal{ H}_0}\right\}}\geq\gamma
\end{equation}
%
%
where $p\left\{ { T_1}, { T_2}, ..., { T_M|\mathcal{ H}_j} \right\}$  is the joint probability distribution of local soft decisions under the $j^{th}$ hypothesis. However, $T_i$ has a $\chi^2$  distribution under $\mathcal{H}_0$ and a non-central $\chi^2$ under $\mathcal{H}_1$, which means evaluation of the LRT in (\ref{eq:LRT}) is complicated. Consequently, we evoke the central limit theorem to simplify the distribution of $T_i$ when $N$ is sufficiently large. So the distribution of any  $T_i$ can be adequately approximated by a Gaussian distribution with the following mean and variance:
%
%
\begin{eqnarray}
\hspace{-0.33cm}\label{eq5554}
 \mathrm{E}\left\{T_i|\mathcal{ H}_0\right\} \hspace{-0.1cm}&\hspace{-0.1cm}=\hspace{-0.1cm}& N\sigma_i^{2}, \ \  \ \ \ \mathrm{Var}\left\{T_i|\mathcal{ H}_0\right\}=2N\sigma_i^{4} \label{eq:T_i-mean}  \\ 
\mathrm{E}\left\{T_i|\mathcal{ H}_1\right\}\hspace{-0.1cm} &\hspace{-0.1cm}=\hspace{-0.1cm}& \hspace{-0.1cm}N\sigma_i^{2}\left(1+\xi_i \right),  \mathrm{Var}\left\{T_i|\mathcal{ H}_1\right\}\hspace{-0.1cm}=\hspace{-0.1cm}2N\sigma_i^{4}\left(1\hspace{-0.1cm}+\hspace{-0.1cm}2\xi_i \right)\nonumber  \\
\hspace*{-0.25cm}
\vspace{-0.3cm}
\end{eqnarray}
%
%
where $\xi_i =\sum \limits_{n=1}^N  s_i^{2}\left(n\right)/N\sigma_i^{2}$ is the SNR at the $i^{th}$ SN. \\

Since the noise at different SNs is independent, it can easily be shown \cite{les8558} that the log-likelihood ratio test (LLR) takes the form
%
%
\begin{equation}\label{eq:LLR}
T_f=\sum \limits_{i=1}^M\left(\frac{\left(  T_i-N\sigma_i^{2}\right)^2}{2N\sigma_i^{4}}-\frac{\left(  T_i-N\sigma_i^{2}\left(1+\xi_i \right)\right)^2}{2N\sigma_i^{4}\left(1+2\xi_i \right)}\right)\geq\gamma'  
\end{equation}
%
%
where $\gamma' =2\ln\left(\prod\limits_{i=1}^M\gamma\left(\frac{\sqrt{2N\sigma_i^{4}}}{\sqrt{2N\sigma_i^{4}\left(1+2\xi_i \right)}}\right)\right)$ .

The LLR can be further simplified by completing the square in (\ref{eq:LLR})  to yield
%
%
\vspace{-0.2cm}\label{www}
\begin{eqnarray}
T_f &=& \sum \limits_{i=1}^Ma_i\left(T_i-b_i\right)^2  \label{eq:OFR} \\
a_i  &=& \frac{\xi_i}{N\sigma_i^{4}\left(1+2\xi_i \right)} \label{eq:a_i} \\
b_i &=& \frac{N\sigma_i^{2}}{2} \label{eq:b_i}.
\end{eqnarray}
%
%

The fusion rule in (\ref{eq:OFR}) has an interesting interpretation. It is, in fact, the \emph{weighted distance} in the $M$-dimensional space between the local soft test statistic and half of its mean under the null hypothesis (see (\ref{eq:T_i-mean})). It is also clear that SNs with lower noise get more weight in the fusion process. Another interesting note here is that at high SNR $(\xi_i)$ the weight $a_i$ depends only on the noise power at the SN and not on the measured signal energy.   
\subsection{Suboptimal Fusion Rules}

Now since the optimal fusion rule in (\ref{eq:OFR}) requires the exact knowledge of the SNR $(\xi_i)$, it cannot be realized in practice. However, its structure can be used to formulate implementable suboptimal rules. So now we propose three suboptimal rules:  weighted fusion, equal fusion and optimum linear fusion. 
\subsubsection{Weighted and Equal Fusion Rules}

The weighted fusion rule takes the same structure as (\ref{eq:OFR}). However (for large $\xi_i)$ $a_i$ in (\ref{eq:a_i}) is replaced by $a^w_i=1/2N\sigma^4_i$ and we let $b^w_i=b_i$. This rule approaches the optimal one when the SNR is large, as discussed earlier. 

As for the equal fusion rule, equal weight is given for all the SNs, i.e., $a^e_i=1$ for all $i=1, 2,\cdots,M$. Also, $b^e_i=b_i$.

\subsubsection{Optimum Linear Fusion Rule}

Now we examine the (sub-optimal) linear fusion rule:
%
%
\begin{equation}\label{eq10}
     T_f^l=\sum \limits_{i=1}^M \alpha_i  T_i.
\end{equation}
%
%
and the optimal weights to maximize the probability of detection are
%
%
\begin{equation}\label{eq:linear_fusion}
\alpha_i = \frac{\xi_i}{N\sigma^2_i \left( 1 + 2 \xi_i \right) }.
\end{equation}
%
%
\indent Due to the lack of space however, the proof is omitted but a detailed discussion can be founded in \cite{les8558}. However, the above weights are not realizable due to requiring a-priori knowledge of $\xi_i $.

%
%
\section{Quantized Soft Decision Fusion}
\label{sec:Quantized Soft Decision Fusion}

The previous fusion rules assume the availability of an infinite bandwidth to send the exact $T_i$. So let the quantized test statistic ($\hat T_i$) at the $i^{th}$ sensor be modeled (with $L_i$ bits) as
\begin{equation}\label{eq6}
     \hat T_i=T_i+v_i
\end{equation}
where $v_i$\footnote{$v_i$ in (\ref{eq6}) is the quantization noise independent of $w_i\left(n\right)$ (in (1) and (2)) for all $n$ and $i$.}  is the quantization noise with uniform distribution in the interval $ [-B, B]$ and variance
%
%
\begin{equation}\label{eq8}
\sigma_{v_i}^{2}=\frac{B^{2}}{3\times 2^{2L_i}}.
\end{equation}
%
%
\indent However, the distribution of $\hat{T}_i$ can be approximated \cite{les8558} by a Gaussian distribution  with mean and variance:
\begin{equation}\label{eq9}
\begin{aligned}
\hspace{-2.1cm}\mathrm{E}\left\{\hat T_i|\mathcal{H}_0\right\}\hspace{-0.0cm}= \hspace{-0.1cm}N\sigma_i^{2}, \ \mathrm{Var}\left\{\hat T_i|\mathcal{H}_1\right\}\hspace{-0.05cm}= \hspace{-0.05cm} 2N  \sigma_i^{4} \left(1+2\xi_i \right)  \hspace{-0.1cm}+\hspace{-0.05cm}\sigma_{v_i}^{2} \hspace{0.36cm}\\ 
\mathrm{E}\left\{\hat T_i|\mathcal{H}_1\right\}=N\sigma_i^{2}\left(1+\xi_i \right), 
\mathrm{Var}\left\{\hat T_i|\mathcal{H}_0\right\}\hspace{-0.05cm}=\hspace{-0.05cm} 2N \sigma_i^{4}\hspace{-0.05cm}+\hspace{-0.05cm}\sigma_{v_i}^{2}. \hspace{0.3cm}
\end{aligned}
\end{equation}
\subsection{Quantized Optimal/Suboptimal Fusion Rule}

Since the $\hat{T}_i$'s are now Gaussian, then in a similar manner to Section \ref {sec:Soft Decision Fusion Rules}, the log-likelihood ratio test with quantization can be shown to be 
%
%
\begin{equation}\label{eq336}
T_f^q=\sum \limits_{i=1}^M\hspace{-0.1cm}\left(\hspace{-0.1cm}\frac{\left( \hat T_i-N\sigma_i^{2}\right)^2}{2N\sigma_i^{4}+\sigma_{v_i}^{2}}-\frac{\left( \hat T_i-N\sigma_i^{2}\left(1+\xi_i \right)\right)^2}{2N\sigma_i^{4}\left(1+2\xi_i \right)+\sigma_{v_i}^{2}}\hspace{-0.05cm}\right)\hspace{-0.1cm}\geq \hspace{-0.1cm}\gamma''  
\end{equation}
%
%
where $\gamma'' =2\ln\left(\prod\limits_{i=1}^M\gamma\left(\frac{\sqrt{2N\sigma_i^{4}+\sigma_{v_i}^{2}}}{\sqrt{2N\sigma_i^{4}\left(1+2\xi_i \right)+\sigma_{v_i}^{2}}}\right)\right)$ .
As before, (\ref{eq336}) can be now written in the following form
%
%
\begin{eqnarray}\label{eq338}
T^q_f &=& \sum \limits_{i=1}^Ma^q_i\left(\hat T_i-b^q_i\right)^2 \\
a^q_i &=& \frac{\xi_i}{N\sigma_i^{4}\left(1+2\xi_i +\frac{\sigma_{v_i}^{2}}{2N\sigma^4_i}\right) \left( 1 + \frac{\sigma_{v_i}^{2}}{2N\sigma^4_i} \right)  } \\
 b^q_i &=& \frac{N\sigma_i^{2}}{2 }-\frac{\sigma_{v_i}^{2}}{4\sigma_i^{2}}.
\end{eqnarray}
%
%
\indent Note that $T^q_f \rightarrow T_f$ as $\sigma^2_{v_i}\rightarrow 0$ for all $i$. Consequently, $a^q_i \rightarrow a_i$ and $b^q_i \rightarrow b_i$ under the previous condition as well. More interestingly however, is that $T^q_f \rightarrow T_f$ as $N \rightarrow \infty$, regardless of $\sigma^2_{v_i}$. This implies that bandwidth can be saved but at the expense of increasing both the number of collected measurements and also the detection delay.


As for the suboptimal (quantized) fusion rule, it can be easily shown that
%
%
\begin{equation}
 a^{wq}_i = \frac{1}{ N\sigma^4_i  \left( 1 +\frac{\sigma^2_{v_i}}{2N\sigma^4_i}  \right)^2 } \label{eq:a^wq}
\end{equation}
%
%
$a^{eq}=1$ and $b^{eq}=b^{wq}=b^q_i$. 

\subsection{Quantized Optimal Linear Fusion Rules}

The quantized version of the linear fusion weights in (\ref{eq:linear_fusion}) can be shown to be \cite{les8558}
%
%
\begin{equation}\label{eq3.111}
\alpha^q_i =  \frac{ \xi_i}{2\sigma^2_i\left[1+2\xi_i+\frac{\sigma^{2}_{v_i}}{N \sigma_i^{2}}\right]}.
\end{equation}
%
%
\indent If the SNR is large, i.e., when either $\sigma^2_{v_i} \rightarrow 0$  or $N \rightarrow \infty$   then it follows that $\alpha^q_i  \rightarrow \alpha_i $.
%
%
\section{Optimum sensor transmit power allocation}
\label{sec:Optimum sensor transmit power allocation}

The performance of the proposed quantized fusion rules approach the performance of their unquantized  counterparts if the number of (test statistic) bits is sufficiently large. However, this entails a large transmission power as predicted by (\ref{eq:capacity}). So, we desire to strike a trade-off between the fusion rule's performance and transmit power. To this end, we first need to adopt an optimization criterion. A natural one is the probability of detection, which depends on the distribution of the fusion rule. So letting $U_i=\left(\hat {T_i}-b_i\right)^2$ then the optimum fusion rule can be written as
%
\begin{equation}\label{eq45}
T_f^q=\sum \limits_{i=1}^Ma_i^qU_i . 
\end{equation}
%
%
The mean and variance of $U_i$ under $\mathcal{H}_0$ and $\mathcal{H}_1$  are now given in (\ref{eq1237}).
%
%
\begin{figure*}[h!tb]
\small
\begin{eqnarray*}
\hspace{+0.18cm}\mathrm{E}\left\{U_i|\mathcal{H}_0\right\}=2N\sigma_i^{4}+N^{2}\sigma_i^{4}+\sigma_{v_i}^{2}-2b_iN\sigma_i^{2}+{b_i}^{2}, \  \ \mathrm{Var}\left\{U_i|\mathcal{H}_0\right\}=\mathrm{Var}\left\{\hat T_i|\mathcal{H}_0\right\}\left[4N^2\sigma_i^{4}+2\mathrm{Var}\left\{\hat T_i|\mathcal{H}_0\right\}+4b_i^2 -8N b_i \sigma_i^{2}\right]
\end{eqnarray*}
\vspace{-0.75cm}
\begin{eqnarray*}\
\hspace{-7.7cm}\mathrm{E}\left\{U_i|\mathcal{H}_1\right\}\hspace{-0.07cm}=\hspace{-0.1cm}\mathrm{E}\left\{\hat T_i|\mathcal{H}_1\right\}^2+\mathrm{Var}\left\{\hat T_i|\mathcal{H}_1\right\}-2b_i\left(N\sigma_i^{2}+N \sigma_i^{2}\xi_i\right)+b_i^{2}
\end{eqnarray*}
\vspace{-0.75cm}
\begin{eqnarray}\label{eq1237}
\hspace{-2cm}\mathrm{Var}\left\{U_i|\mathcal{H}_1\right\}\hspace{-0.1cm}=\hspace{-0.1cm}4\mathrm{E}\left\{\hat T_i|\mathcal{H}_1\right\}^2\mathrm{Var}\left\{\hat T_i|\mathcal{H}_1\right\}+2\mathrm{Var}\left\{\hat T_i|\mathcal{H}_1\right\}^2  + 4b_i^{2}\mathrm{Var}\left\{\hat T_i|\mathcal{H}_1\right\}-8b_i\mathrm{E}\left\{\hat T_i|\mathcal{H}_1\right\}\mathrm{Var}\left\{\hat T_i|\mathcal{H}_1\right\}
\end{eqnarray}
\hrulefill
\end{figure*}
%
%
%
 Using the central limit theorem, $T_f^q$ can be approximated by a Gaussian distribution
%
%
\begin{equation}\label{eq9}
T_f^q\sim\left\{
\begin{aligned}
\mathcal{N}\left(\mathrm{E}\left\{T_f^q|\mathcal{H}_0\right\},\mathrm{Var}\left\{T_f^q|\mathcal{H}_0\right\}\right) \mathrm{under} \ \mathcal{H}_0\\
\mathcal{N}\left(\mathrm{E}\left\{T_f^q|\mathcal{H}_1\right\},\mathrm{Var}\left\{T_f^q|\mathcal{H}_1\right\}\right) \mathrm{under} \ \mathcal{H}_1 
\end{aligned}
\right.
\end{equation}
%
%
where
%
%
\begin{equation}\label{eq9}
\begin{aligned}
\mathrm{E}\left\{T_f^q|\mathcal{H}_0\right\}=\sum \limits_{i=1}^M a_i^q \mathrm{E}\left\{U_i|\mathcal{H}_0\right\} \hspace{0.03cm}\ \ \ \ \ \ \ \ \ \ \ \ \ \  \ \\ 
\mathrm{E}\left\{T_f^q|\mathcal{H}_1\right\}=\sum \limits_{i=1}^M a_i^q \mathrm{E}\left\{U_i|\mathcal{H}_1\right\}\ \ \ \ \ \ \ \ \ \ \ \ \  \  \  \\
\mathrm{Var}\left\{T_f^q|\mathcal{H}_0\right\}=\sum \limits_{i=1}^M \left(a_i ^q\right)^2\mathrm{Var}\left\{U_i|\mathcal{H}_0\right\} \ \ \ \ \ \ \\
 \mathrm{Var}\left\{T_f^q|\mathcal{H}_1\right\}=\sum \limits_{i=1}^M \left(a_i ^q\right)^2\mathrm{Var}\left\{U_i|\mathcal{H}_1\right\}. \ \ \ \
\end{aligned}
\end{equation}
%
%

It can be readily shown that the detection probability as a function of the false alarm probability has the form 
%
%
\begin{equation}\label{eq2444}
P_d = Q
\left(\frac{Q^{-1}\left(P_{fa}\right) \sqrt {\mathrm{Var} \left\{T_f^q|\mathcal{H}_0\right\}} - \Psi}{\sqrt {\mathrm{Var}\left\{T_f^q|\mathcal{H}_1\right\}}}\right)
\end{equation}
%
%
where $Q(\cdot)$ is the $Q$-function and $\Psi=\mathrm{E}\left\{T_f^q|\mathcal{H}_1\right\} -\mathrm{E}\left\{T_f^q|\mathcal{H}_0\right\}$. The probability of detection implicitly depends on the transmission power through the relationships  (\ref{eq1237}) and (\ref{eq9}). Based on this, we can optimize the transmission powers ($p_i$) to maximize $P_d$  under the constraint of a maximum aggregate transmit power budget ($P_t$):
%
\begin{equation}\label{eq20}
\begin{aligned}
\hspace{0.3cm} \boldsymbol p_{opt} =\  \underset{\hspace{0.6cm}  \boldsymbol p}{\text{arg \ max}}
\hspace{0.08cm}{P_d}\left(\boldsymbol p \right) \ \ \ \ \ \ \ \ \ \ \  \ \ \ \ \ \ \ \ \ \ \ \  \  \ \ \ \ \ \ \ \ \ \ \  \  \ \ \ \ \\\ \quad
\text{\hspace{-2cm}subject to} 
 \sum \limits_{i=1}^M p_i \leq P_t \ \mathrm{for} \  \ p_i\geq0,  \ i = 1, \ldots, M \hspace{1.28cm}
\end{aligned}
\end{equation}
\noindent where $\boldsymbol p=[p_1, p_2, \ldots, p_M].$
%
%
Now  ($\ref{eq20}$) is difficult to solve and there is no closed form solution. Hence,  we propose a numerical solution by adopting the spatial branch-and-bound strategy \cite{les8343} using the YALMIP optimization tools \cite{les8355}.  In the first step of the algorithm, we start by applying a standard  nonlinear solver to obtain a locally optimal solution and then set it as an upper bound on the achievable objective. Secondly, in each node, a convex relaxation of the model is derived, and the resulting convex optimization problem is solved. We then assign this as a lower bound. Bound tightening using \cite{les8355} is applied iteratively to  detect and eliminate redundant constraints and variables, and tighten the bounds where possible. The algorithm outline is summarized in ${Algorithm1}$\cite{les8558}. \\
\indent The aim of the algorithm is to obtain the global minimum of the function $\beta\left(\boldsymbol p\right)=\frac{Q^{-1}\left(P_{fa}\right) \sqrt {\mathrm{Var} \left\{T_f^q|H_0\right\}} - \Psi}{\sqrt {\mathrm{Var}\left\{T_f^q|H_1\right\}}}$ over the solution space $\wp_{start}$ where $\boldsymbol p\in \wp_{start}$. For any $  \wp \subseteq \wp_{start}$ we define $F_{lb}$ (the upper bound) and $F_{ub}$ (the lower bound) as functions that satisfy: $F_{lb}\left( \wp\right)\leq F_{min}\left( \wp\right)\leq F_{ub}\left( \wp\right)$. Then, the global optimum solution $\beta^*=F_{min}\left(\wp_{start}\right)=\mathrm{inf}_{\boldsymbol p\in \wp_{start}}\beta\left(\boldsymbol p\right)$.

%

%
\section{Simulation  Results}
\label{sec:Simulation  Results}
\begin{figure}[htp!]
\vspace{-2.9cm}
    \centerline{\includegraphics[width=90mm ,height=110mm]{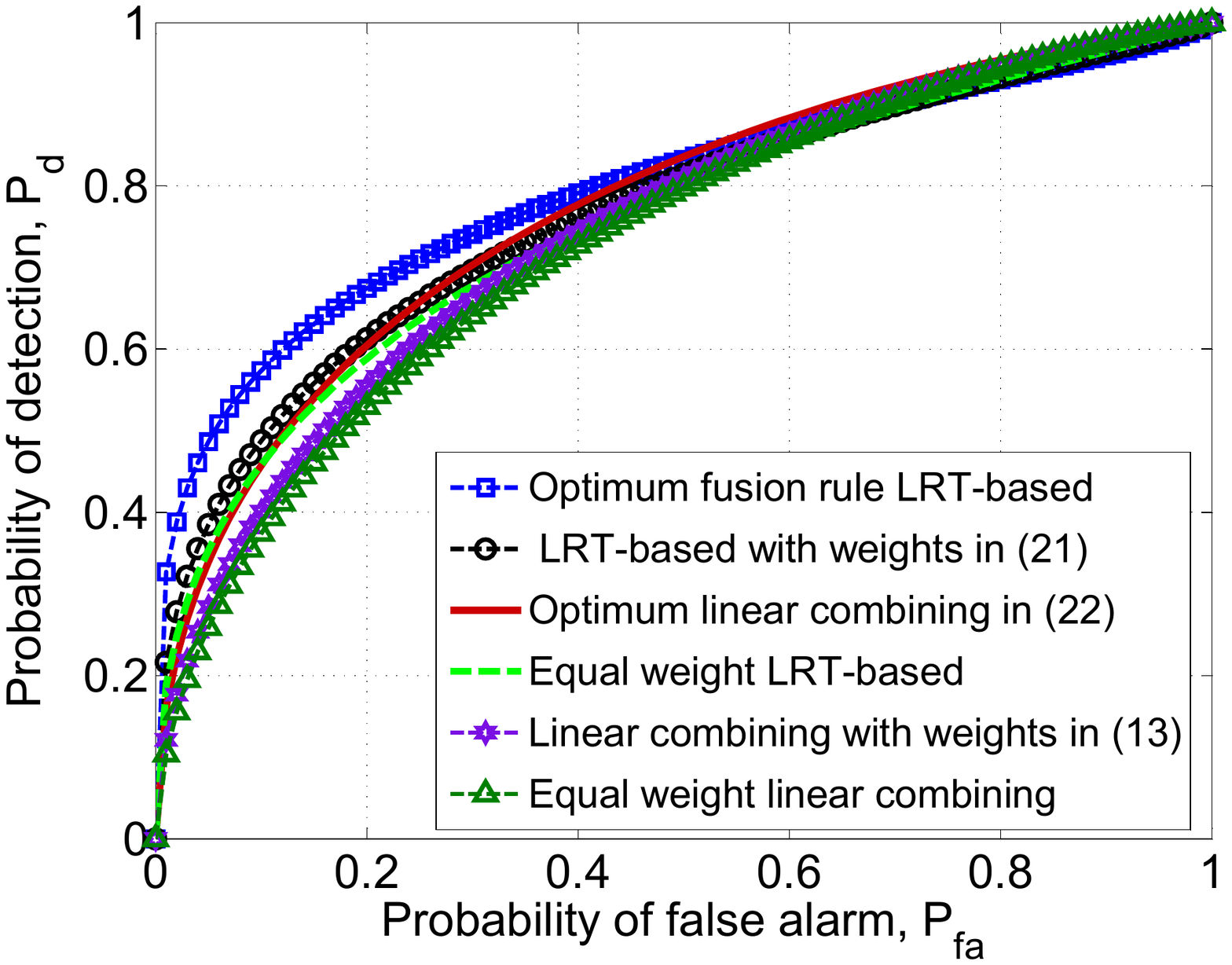}}
\vspace{-2.2cm}
    \caption[Optimum fusion rule and optimum linear combining for $N=10$ samples and $M=10$ sensors.]{\label{fig:llr_1}
    \small Receiver operating characteristics of six different fusion rules for $N=10$, $M=10$, $\xi_a=-8.5$ dB and $B=0.5$.}
\vspace{-0.0cm}
\end{figure}

We simulate a WSN of $M$ SNs detecting an intruder with $s_i(n)=A$, where $A=0.1$. The communication noise variances are arbitrarily set to $\zeta_i=0.1$ for all $i=1, 2,\cdots,M$ (for simplicity). The measurement noise variances are generated randomly and used throughout all the simulations. The average measurement SNR for the network is defined as $\xi_a=10\log_{10}\left(\frac{1}{M}\sum \limits_{i=1}^M \xi_i \right)$. In all simulations we assume perfect knowledge of $\xi_i$.   Fig. \ref{fig:llr_1} shows the receiver operating characteristic for six different fusion rules. It is clear that the optimal fusion rule attains the best performance for ($\xi_a=-8.5$ dB) whereas the worst performance is that of the equal weight linear combining rule. However, all the rules converge when the $P_{fa}$ increases.   
\begin{figure}[htp!]
\vspace{-3.2cm}
\hspace{0cm}
   \centerline{\includegraphics[width=90mm ,height=110mm]{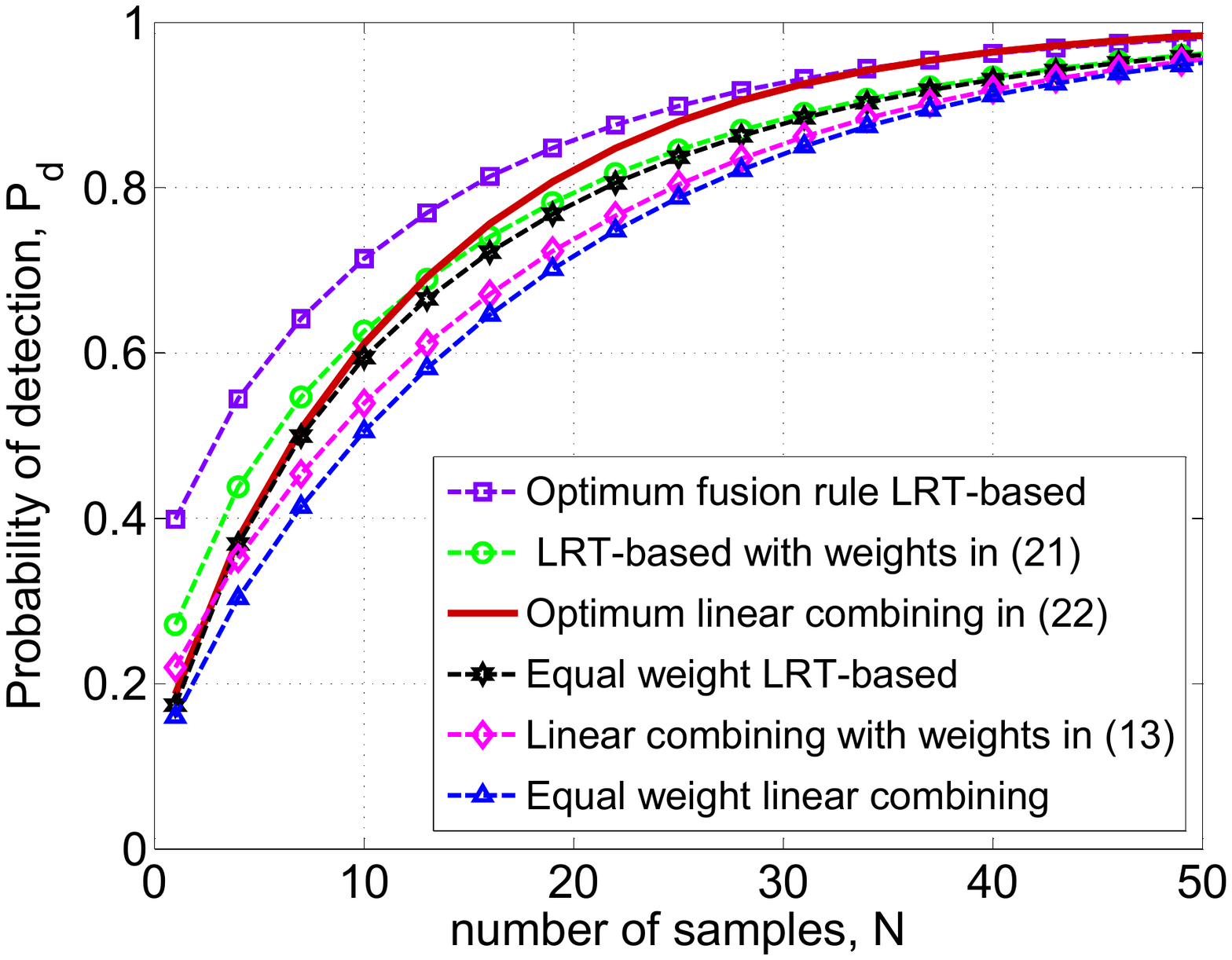}}
\vspace{-2.5cm}
    \caption[Optimum fusion rule and optimum linear combining with $N=10$ samples.]{\label{fig:llr_2}
\small Probability of detection ($P_d$) versus the number of samples ($N$) with $M=20$, $P_{fa}=0.1$, $B=0.5$ and $\xi_a=-8.5$ dB.}
\end{figure}
\begin{figure}[htp!]
\vspace{-2.9cm}
\hspace{0cm}
    \hspace{0.12cm}\centerline{\includegraphics[width=90mm ,height=110mm]{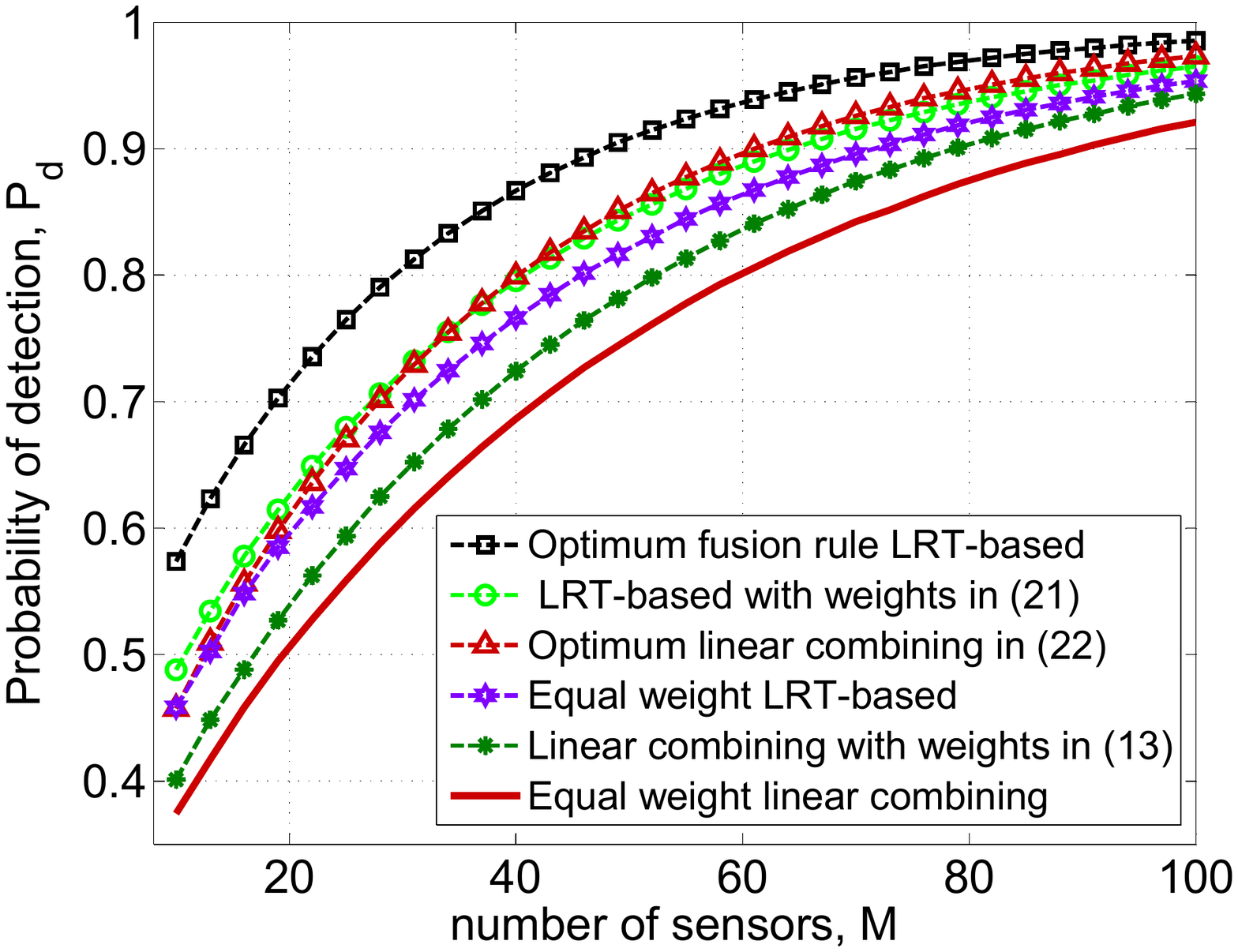}}
\vspace{-2.3cm}
    \caption[Receiver operating characteristic for 10 samples.]{\label{fig:llr_3}
\vspace{-0cm}
  \small Probability of detection ($P_d$) versus number of sensors ($M$) for $N=10$, $P_{fa}=0.1$, $\xi_a=-8.5$ dB and $B=0.5$.}
\end{figure}
\begin{figure}[htp!]
\vspace{-3.3cm}
\hspace{0cm}
    \hspace{0.12cm}\centerline{\includegraphics[width=90mm ,height=113mm]{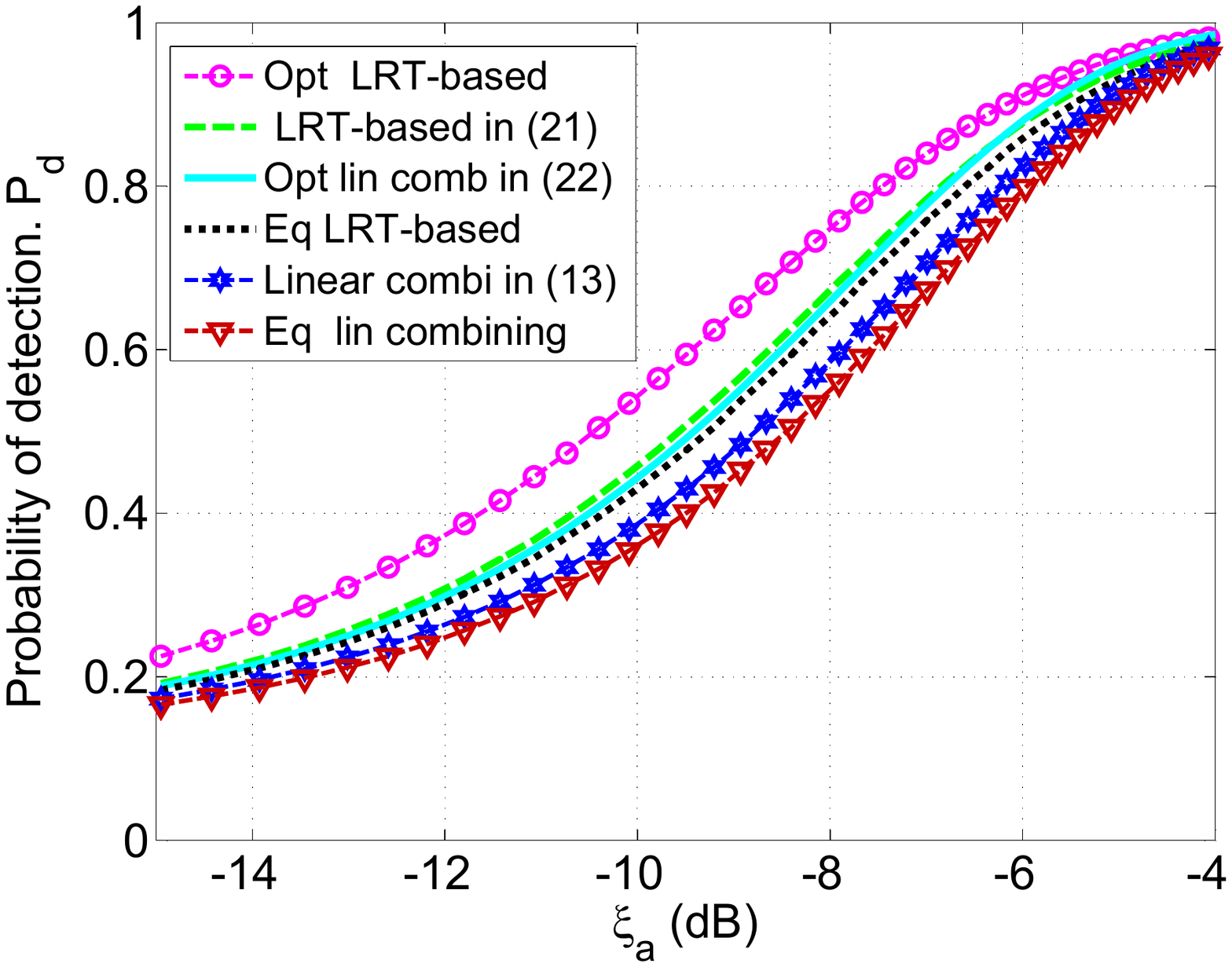}}
\vspace{-2.4cm}
    \caption[Receiver operating characteristic for 10 samples and M=10 sensors.]{\label{fig:SNR}
\vspace{-0.0cm}
   \small Probability of detection ($P_d$) versus the signal to noise ratio ($\xi_a$) for $M=20$, $N=10$, $P_{fa}=0.1$ and $B=0.5$.}
\end{figure}
\begin{figure}[htp!]
\vspace{-3.2cm}
\hspace{0cm}
    \hspace{0.12cm}\centerline{\includegraphics[width=90mm ,height=113mm]{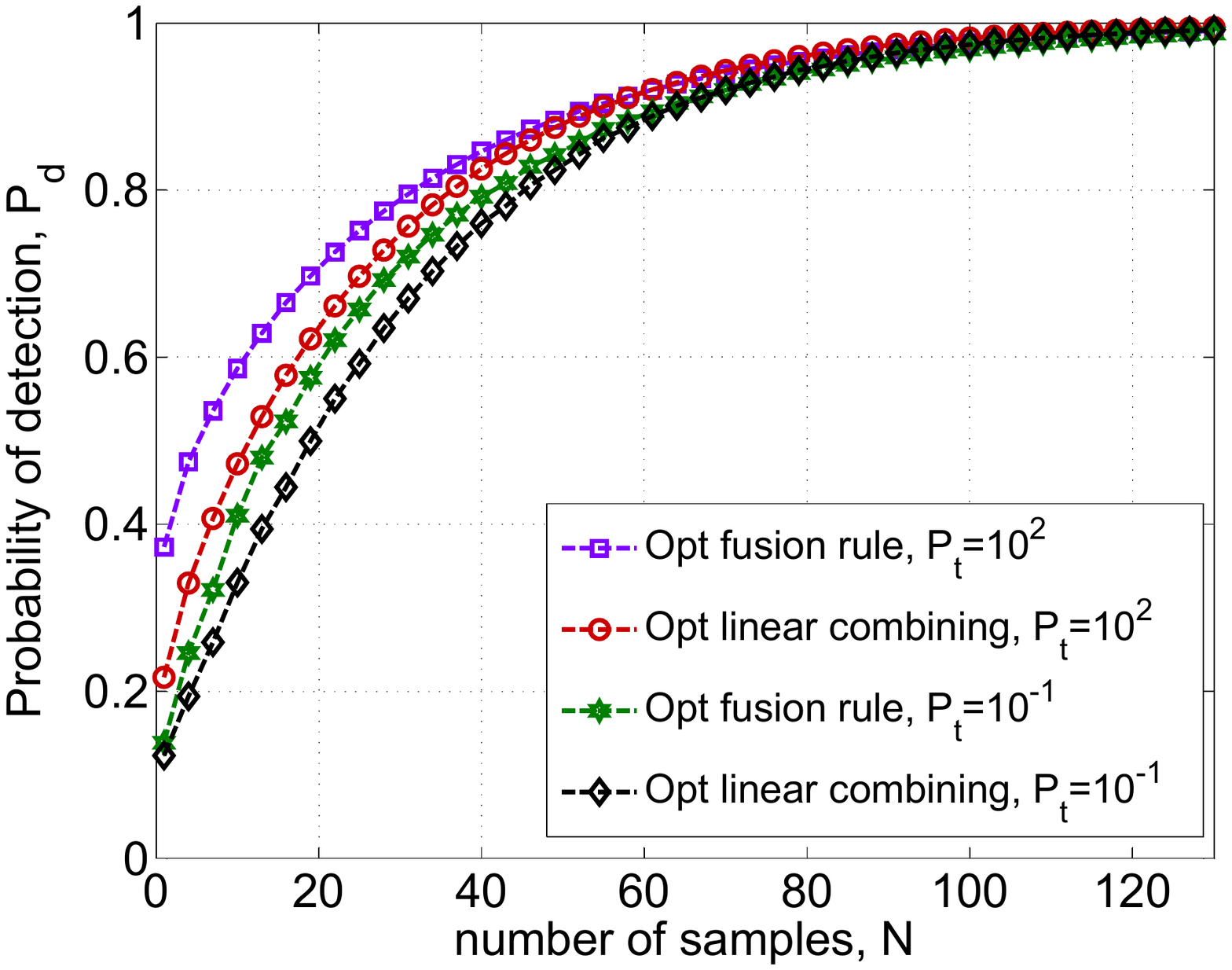}}
\vspace{-2.6cm}
    \caption[Receiver operating characteristic for 10 samples and M=10 sensors.]{\label{fig:llr_4}
\vspace{-0.0cm}
  \small  Probability of detection ($P_d$) versus the number of samples ($N$) for $M=10$ sensors, $P_{fa}=0.1$, $\xi_a=-8.5$ dB and $B=1$.}
\end{figure}
\begin{figure}[htp!]
\vspace{-4.6cm}
\hspace{0cm}
    \hspace{0.12cm}\centerline{\includegraphics[width=90mm ,height=150mm]{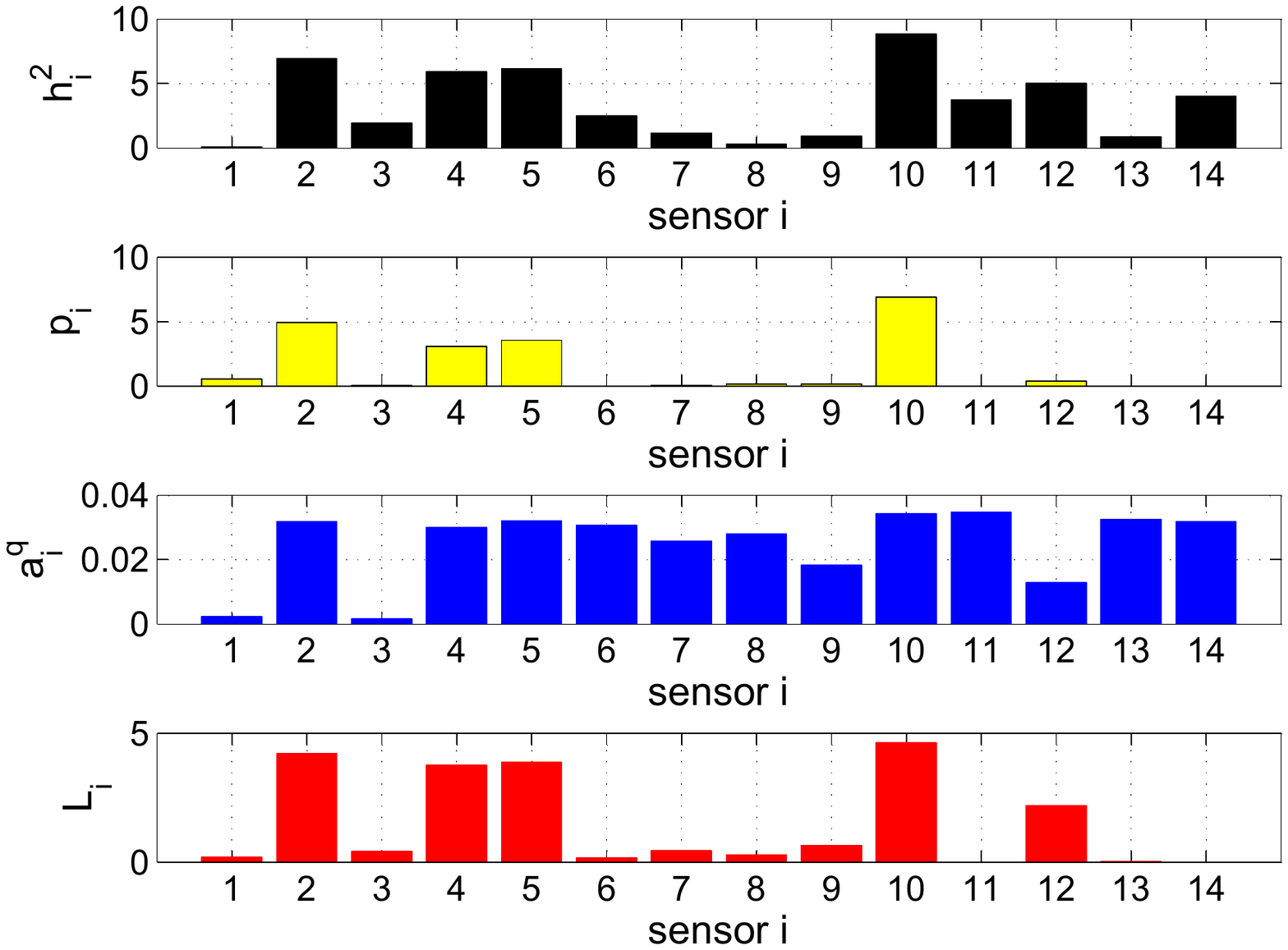}}
\vspace{-3.8cm}
    \caption[Receiver operating characteristic for 10 samples and M=10 sensors.]{\label{fig:llr_7}
\vspace{-0.0cm}
 \small  Optimum sensor transmit power and channel quantization  bits allocation for $N=10$, $P_{fa}=0.1$, $\xi_a=-8.5$ dB and $P_t=20$.}
\end{figure}
In Fig. \ref{fig:llr_2}  the effect of the number of measurement samples ($N$) on $P_d$ is shown at a fixed $P_{fa}$. Obviously, as $N$ increases $P_d$ improves for all algorithms. Interestingly, the optimal linear fusion rule outperforms the suboptimal LRT-based one. This is explained by the structure of (21) where for large (but finite) $N$ the effect of  $\sigma^2_{v_i}$ (quantization noise variance) is still noticeable. A similar trend is noticed in Fig.\ref{fig:llr_3}, in which $P_d$ is plotted against the number of SNs, ($M$), for a fixed $N$. The $P_d$ performance of both LRT-based and linear combining schemes as a function of the average SNR ($\xi_a$) is shown in Fig. \ref{fig:SNR}. Fig. \ref{fig:llr_4} on the other hand, exhibits the effect of the transmission power $p_i$  on $P_d$. Increasing $p_i$ leads to a larger number of allocated bits, through  (\ref{eq:capacity}), and consequently less quantization variance, which ultimately  improves the detection performance. Interestingly, the dependence of $P_d$ on $p_i$ is alleviated when $N$ is increased, since the effect of the quantization noise is mitigated as predicted by (19) and (\ref{eq3.111}). In Fig. \ref{fig:llr_7}, we report the optimized sensor transmit power and the corresponding number of bits allocated to quantize $T_i$ by applying the branch and bound algorithm \cite{les8343}. Clearly we allocate more power and bits to the best channels. However, note that the power and bit allocation is also affected by the weights $a_i^q$ in (19) which are a function of the signal to noise ratio $\xi_i$. For instance, consider sensor 12 which has a relatively good channel gain, but the corresponding local $\xi_i$ is bad. Hence, it will allocate a relatively small amount of the transmit power. Those SNs with bad channels are allocated \emph{zero} bits, i.e., they will be censored or prevented from transmission.
%
%

\section{CONCLUSION}
\label{sec:conclusion}
We have shown that the optimal fusion (see (9)) for energy-based soft decisions is actually the weighted distance of the decisions from their mean under the null hypothesis. Realizable suboptimal fusion rules derived from the optimal one are proposed as well, in which more weight in the actual fusion are given to decisions with better sensing quality. We show that the effect of quantization on the detection performance can be mitigated by increasing the number of measurements ($N$), or equivalently incurring more delay in the system. Finally, the SN's  transmission power has been optimally allocated. Intuitively, more power is given to SNs having better channel gains and consequently increased number of bits.


\end{document}